  \documentclass{vgtc}                          
\let\ifpdf\relax

\usepackage[hyphens]{url}
\usepackage{mathptmx}
\usepackage{graphicx}
\usepackage{times}
\usepackage{subfigure}
\usepackage{color}
\usepackage{verbatimbox}
\newcommand{\specialcell}[2][c]{%
  \begin{tabular}[#1]{@{}c@{}}#2\end{tabular}}

\vgtcinsertpkg

\title{Navigable Videos for Presenting Scientific Data\\ on Head-Mounted Displays}

\author{Jacqueline Chu\thanks{e-mail: sjchu@ucdavis.edu}\\ %
        \scriptsize University of California, Davis %
\and Leonardo Ferrer\thanks{e-mail:lferrer@ucdavis.edu}\\ %
     \scriptsize University of California, Davis %
\and Min Shih\thanks{e-mail:minshih@ucdavis.edu}\\ %
     \scriptsize University of California, Davis %
\and Kwan-Liu Ma\thanks{e-mail:ma@cs.ucdavis.edu}\\ %
    \scriptsize University of California, Davis %
    }

\abstract{
Immersive, stereoscopic viewing enables scientists to better analyze the spatial structures of visualized physical phenomena.  However, their findings cannot be properly presented in traditional media, which lack these core attributes.  Creating a presentation tool that captures this environment poses unique challenges, namely related to poor viewing accessibility.  Immersive scientific renderings often require high-end equipment, which can be impractical to obtain.  We address these challenges with our authoring tool and navigational interface, which is designed for affordable head-mounted displays.  With the authoring tool, scientists can show salient data features as connected 360$^{\circ}$ video paths, resulting in a ``choose-your-own-adventure'' experience.  Our navigational interface features bidirectional video playback for added viewing control when users traverse the tailor-made content.  We evaluate our system's benefits by authoring case studies on several data sets and conducting a usability study on the navigational interface's design.  In summary, our approach provides scientists an immersive medium to visually present their research to the intended audience--spanning from students to colleagues--on affordable virtual reality headsets.
}

\begin{document}


\firstsection{Introduction}

\maketitle

Scientific studies are often about understanding complex 3D phenomena and structures.  Visualization is a powerful way to present and perceive such information.  By adding stereoscopic effects, the perception of 3D shape, structure, and relationships is enhanced~\cite{laha2012effects}.  However, the technology required for these studies is often expensive, making it difficult for scientists to present their findings in context to an immersive environment.  Immersive, stereoscopic viewing can be a key factor to a scientist's discovery and should not be omitted when sharing these findings.  Since traditional media cannot visually capture these characteristics, we developed a system that produces 360$^{\circ}$ navigable videos: a novel presentation medium designed for viewing on affordable virtual reality (VR) headsets.

However, porting interactive scientific visualization to a VR setting is non-trivial.  Scientific data is inherently large, requires high-precision, and instead of 3D meshes, is often viewed as volumetric structures.  Rendering scientific data relies on expensive rendering algorithms, in which the resulting frame rate and latency are often unacceptable for VR environments~\cite{van2000immersive}.  In this paper, we define \emph{latency} as the time between when the user moves their head to when the image is updated. This definition of latency is oftentimes referred to as \emph{motion-to-photon}~\cite{kanter2015graphics}.  In favor of interactive and high-quality visuals, expensive and specialized hardware can be used, but may not be the most viable solution, especially when the target audience does not have access to that level of equipment.  A more practical solution for porting scientific visualization usually involves some compromise: data compression and sub-sampling, low rendering quality~\cite{funkhouser1993adaptive}, or limited interactions~\cite{noguera2016mobile}.  Sacrificing the visualization's quality is not an ideal solution, as it can be detrimental to how well the content is perceived by the audience.

With these considerations in mind, we circumvent these limitations by developing tools that facilitate the authoring and viewing of interactive and immersive videos of scientific data. Our work focuses on providing a presentation medium for scientists, or \emph{authors}, to present their findings and teachings to their intended audience, which can range from colleagues to students or even to the general public.  Our process is two-fold, consisting of an authoring stage and navigational interface.  Specifically, the authoring stage produces video content that maintains the rendering quality and is in the form of a virtual tour.  The author can preset navigable paths, which form a \emph{roadmap} of connected videos for viewers to ``traverse.''   This evokes a ``choose-your-own-adventure'' style of storytelling when viewing scientific data, giving users adequate control over their viewing experience. In this paper, we designed the navigational interface to be experienced on head-mounted displays (HMDs). With head-tracking support, HMDs allow users to interactively change their viewing direction across 360$^{\circ}$.  Since our authoring tool generates videos that are comprised of a series of panoramic images, the user can view various portions of this scene as if they were fully-enclosed at the center of a spherical display.
To have the content more accessible in typical learning environments, such as classrooms, we targeted affordable HMDs instead of high-end headsets.  These headsets leverage a smartphone's gyroscope and mobile GPU for head tracking and rendering, respectively.  It is a low-cost alternative in comparison to comprehensive VR setups which include built-in motion tracking and require a powerful desktop for rendering.

We believe that our presentation medium is effective for communication purposes such as knowledge sharing or education.  Since scientific data typically involves many variables of interest to manipulate and analyze, the exploration space around the data is highly-dimensional. The tasks required to filter and navigate this high-dimensional interaction space are likely to be complex and overwhelming for inexperienced users such as students.  To this end, our system presents the data in a user-friendly way, since the author predetermines the parameters for viewing and uses their expertise to highlight salient features of the data.  Despite its limited explorability, the video content closely matches the author's intent.  In this work, we define \emph{intent} to be the story or message the scientist would like to convey when presenting their findings to their target audience.  For example, an author can reconstruct the steps of their scientific analysis by creating paths of the dimensional changes that led to their discovery.  By reducing the high-dimensional space of limitless data changes and interactivity possibilities, the author can create a more focused subset of data exploration for the viewer to navigate through.

Succinctly, our work makes the following contributions to scientific visualization:

\begin{itemize}
\itemsep-0.10em
\item{An authoring tool that generates a series of video paths, which reflect the author's intent of refining the interaction space to best present the content to their target audience.}

\item{A navigational interface that supports bidirectional video playback to complement the tailor-made experience from the authoring stage.}

\item{A presentation medium that leverages stereoscopic, 360$^{\circ}$ viewing on affordable HMDs for communication and knowledge-sharing purposes.}

\end{itemize}

By conducting several case studies and a usability study on the viewing experience, we illustrate the benefits of our system when creating immersive, but portable scientific visualizations for accessible presentation.  To better showcase the system's features and results, a video is included in our supplementary materials.  From our results, we believe that our approach can yield similar presentation benefits for other types of visualizations that may want to leverage immersive, stereoscopic viewing, but are too computationally demanding to be rendered in real-time.

\section{Related Work}
Our work encompasses interactive videos, immersive scientific visualization, and animation for the use of storytelling.  Using these concepts together, our system addresses the lack of effective presentation media for scientists to share their research on low-cost immersive, stereoscopic viewing displays.

\subsection{Interactive Video}
In our approach, ``interactive'' refers to providing users control over their navigation of the video paths and 360$^{\circ}$ viewing.  However, what constitutes as an interactive video can be ambiguous~\cite{meixner2012interactive}.  Some features include non-linear playback~\cite{meixner2010siva} and detail-on-demand video summaries~\cite{shipman2003hyper}.  All of these features leverage different forms of interactivity to provide a flexible viewing experience.  Panoramic videos are also considered interactive for when users change their viewing direction during its playback.  Having recognized the benefits of immersion, work has been done to facilitate the production and viewing of immersive videos~\cite{agarwala2005panoramic}.  

In addition, our work includes the authoring of immersive videos that showcase scientific data, similar to that of Stone et al.'s~\cite{stone2016immersive}. In their work, they visualized and produced movies on molecular dynamic simulations involving millions of 3D atoms.  By incorporating omnidirectional, panoramic techniques into their rendering engine, the resulting movies can be viewed on various HMDs. Likewise, we also have added panoramic projection techniques to create immersive and stereoscopic videos; more details are discussed in Section~\ref{section:authoring_tool}.

\subsection{Immersive Scientific Visualization}

Scientific visualization has been shown on a variety of immersive displays: spherical~\cite{amatriain2009allosphere}, large-tiled \cite{reda2013visualizing}, fish-tank~\cite{demiralp2006cave}, CAVEs~\cite{ zhang2001immersive}, and HMDs~\cite{marks2014towards}.  In particular to HMDs, Drouhard et al. proposed design strategies for immersive virtual environments to facilitate the adoption of VR into scientific domains~\cite{drouhard2015immersive}.  They discussed how influential HMDs can be for the scientific community, with one of its key benefits being affordability.  Designed for consumer-available headsets, our system facilitates knowledge sharing, especially in a classroom environment where funding and space are too limited to obtain high-end displays~\cite{roussou2000immersive} like a CAVE.  

To provide a comfortable VR experience, optimization techniques have been developed to improve the viewing and interactive experience around immersive scientific data.  Ebert et al. used a glyph-based volume renderer--which they preferred over isosurface or voxel-based techniques--to provide fast rendering times to support their stereoscopic viewing system~\cite{ebert1996two}.  Kniss et al. implemented a texture-based rendering system for terabyte-sized volume data sets on a high-IO, multi-hardware system~\cite{kniss2001interactive}.  Although only achieving 5 to 10 frames per second, this technique provides low latency by modifying the pipelines' workload, either by rendering small, but multiple portions or data subsampling to render fewer samples per frame.  In recent work, Hanel et al. continuously adjusted the visual quality in favor of stable frame rates and preventing simulation sickness~\cite{hanel2016visual}.  

For our purposes, we found videos and our treatment of them to provide a unique learning and presentation experience.  Exporting to this medium also avoids the side effects that optimization techniques are often associated with, such as the loss of visual quality or network dependencies for rendering.  Our approach preserves the visual quality of advanced rendering techniques and offers interactivity through a roadmap's size and structure of video paths.  Since scientists use their expertise when authoring videos, the resulting navigable videos of scientific data are promising for education, as the application of VR has shown to be useful in other learning domains~\cite{ponder2003immersive, roussou2004learning}.

\subsection{Animation for Storytelling}
In our efforts to support effective content presentation, our system was inspired by storytelling principles.  Scientific visualization have recognized its benefits~\cite{ma2012scientific, wohlfart2006story} and have established frameworks for effective communication to the target audience.  However, these storytelling guidelines and frameworks can be difficult to incorporate in practice. Gershon and Page summarized this challenge as ``a story is worth a thousand pictures'' in which a single static image cannot capture all the multifaceted components of a story~\cite{gershon2001storytelling}.  Fortunately, animation is an effective tool to aid storytelling visualization, but must be applied appropriately to improve user experience and visual discourse~\cite{chevalieranimations}. 

Nowadays, most scientific tool kits include basic animation support for video export.  More comprehensive systems allow changes for a variety of dimensions~\cite{ akiba2010aniviz, limaye2012drishti} by interpolating viewpoint, color mapping, or clipping planes.  Hsu et al. generated animation by automating camera paths from user-specified criteria~\cite{hsu2013multi}, while Liao et al. leveraged a scientist's exploration history~\cite{liao2014storytelling}.  We have found that the underlying models of scientific animations tend to be timelines, which suggest that the animation is linear~\cite{spaniol2006web}.  
 
However, studies have found that users prefer non-linear animations~\cite{lugrin2010exploring, marchetti2015happened}.  Zhang et al. discussed that interactive videos enhance learner-content interactivity which potentially improves learning effectiveness and motivation~\cite{zhang2006instructional}. For reasons like this, our work focuses on the producing and viewing of non-linear stories around the  scientific data.  Similar to Wohlfart and Hauser's work~\cite{wohlfart2007story}, we use a node-link diagram to build non-linear animation.  However, our playback interface is its own stand-alone component and is not integrated into the given renderer.  Instead of allowing user manipulation of the presentation, we limit the user's interaction to navigation of the author's roadmap, which is composed of the fundamental and tailored characteristics of the data.  Having the ability to truly explore can verify a user's understanding~\cite{gratzl2016visual}, but an advantage of our interaction design is that users can stay focused.  If we allowed users to deviate from the intended storyline, they may become distracted~\cite{steele2010beautiful}.

\section{System Overview}

\begin{figure}[h]
\centering
\includegraphics[width=1.0\linewidth]{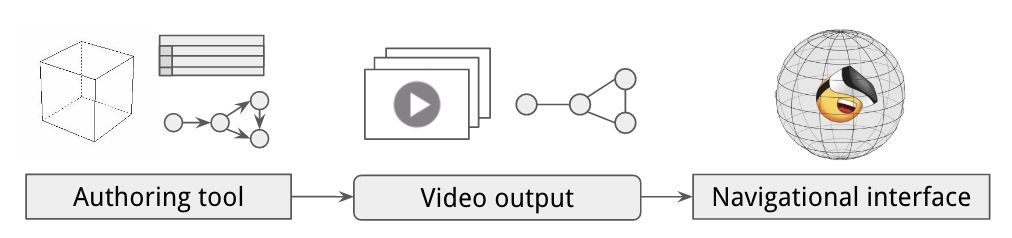}
\caption{An illustrative system overview of our authoring tool and navigational interface.  The video output is an intermediary stage---comprised of videos and a roadmap---between the authoring and navigational components.}
\label{pipeline}
\end{figure}

\begin{figure*}[t]
  \centering
  \includegraphics[width=0.75\linewidth]{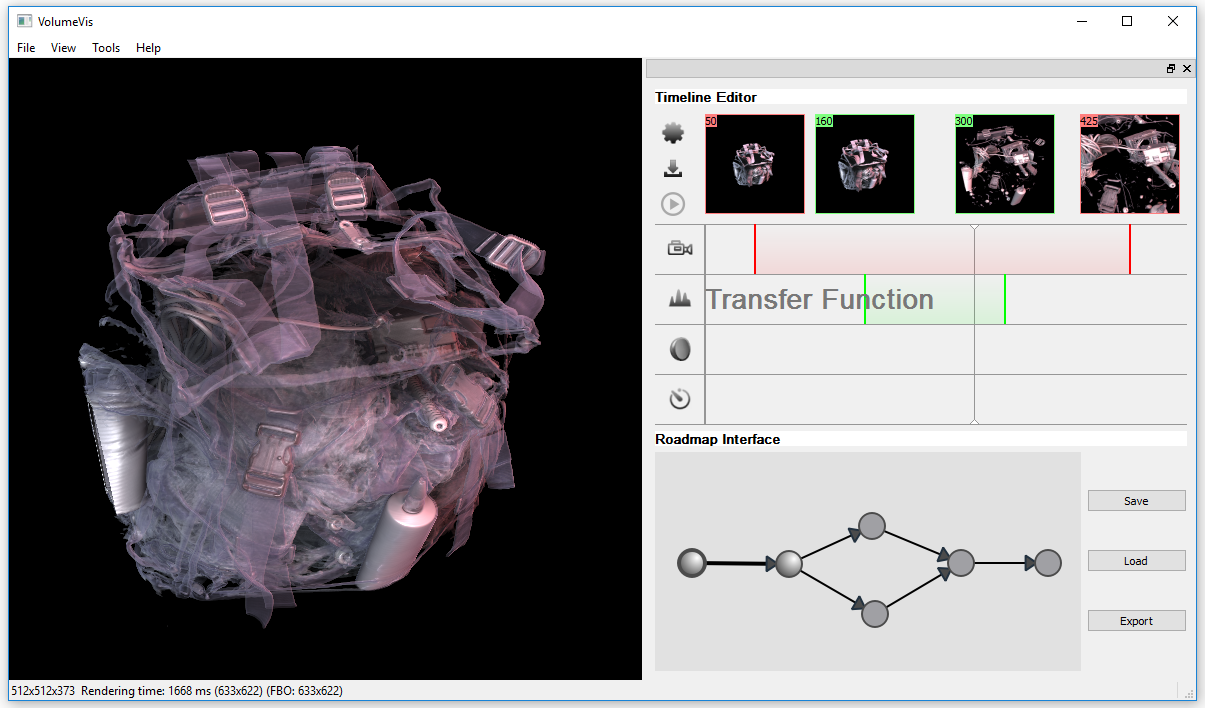}
\caption{An example of using the authoring tool around a backpack data set, which was generated from evaluating nondestructive testing methods.  The authoring tool's two subcomponents, timeline editor and roadmap interface, are shown on the right of our volume renderer.  The timeline editor shows dimensional changes involving viewpoint and transfer function (TF).  Each lane has an icon and its keyframes are color coded to match its respective dimension.  Shown with frame numbers, snapshots are placed above its respective keyframe and reflect the change made.  Options on the top-left help with creation of a single sequence of animation.  The roadmap interface shows the roadmap that connects and structures the non-linear animation.  Nodes with specular highlights store a keyframe, whereas gray ones do not.  The thicker black line represents the current video the author is modifying.  Although video playback and dimension interpolation are bidirectional, edges are displayed to be directed to denote the corresponding video's start and end.  This provides authors a reference on which ends of the video they are modifying and connecting the dimensional changes.  The options on the right can be used to save and load roadmaps to apply the changes to other data sets. Using the export button, all the videos are generated and connected appropriately.}
\label{fig:author_tool}
\end{figure*}

Our system is comprised of two distinct components as shown in Figure~\ref{pipeline}.  The authoring tool enables scientists to construct stories around their data in the form of a roadmap, while the navigational interface facilitates the immersive, stereoscopic viewing of the resulting content on HMDs.  For the authoring and viewing of navigable videos, a typical workflow starts with the author visualizing the data of interest.  Using a renderer that has our authoring tool integrated, a scientist can utilize its subcomponents---the \textit{timeline editor} and \textit{roadmap interface}---to build each video.  For a single animation segment, the timeline editor can interpolate multiple data dimensions for the scientist to highlight key characteristics of the data.  Using the roadmap interface, the author can connect the animation segments to customize how users should experience the content.  Once the content is finalized, the authoring tool exports the roadmap and a series of videos as input for the navigation component.  The roadmap serves as the underlying structure of the video navigation, such that the navigational interface can play the next video by using the viewer's current position in the roadmap.

In the remainder of this section, we detail both the authoring tool and navigational interface.  Since the scope of our work mainly targets HMDs that require smartphones, the details disclosed are in context of mobile hardware.  In this work, we used a Samsung Galaxy S6 and recommend that the use of the navigational interface should be on phones with similar specifications.  For more examples of the system components and resulting videos, please refer to the supplementary video.

\subsection{Authoring Tool}
\label{section:authoring_tool}

This authoring stage can be seen as a preprocessing step to produce high-quality visuals that can be presented comfortably in VR headsets.  This tool is designed to be a modular animation library which can be integrated into different types of rendering systems, such as those that render 3D meshes or non-uniform grid data.  In this paper, we have visualized volumetric data that is uniformly structured on a grid.

This tool was implemented using C++, Qt, and FFMPEG for video export. Although it is designed to be  renderer-agnostic, the authoring tool must be connected to the given renderer using Qt's event framework. If it is not set up with these dependencies, the renderer must be able to export videos that match the specifications described in Section~\ref{subsection:video_output}.  Omnidirectional rendering must also be used to enable stereoscopic viewing, similar to that generated by our camera model, which is described in Section~\ref{subsection:renderer}.  Finally, a roadmap metadata file, which contains additional video information, will then need to be created.  Once it is generated, the video output can then be used as input into our navigation interface.

\subsubsection{Renderer}
\label{subsection:renderer}
An interactive renderer allows scientists to experiment with various rendering parameters, such as color or viewpoint, and produce visuals that effectively showcase the unique data features.  However, a small caveat exists when rendering stereoscopic content: The renderer must provide an image for each eye.  Since scientific visualizations involve expensive rendering algorithms, we have implemented Google's Omni-directional Stereo (ODS) camera model.  ODS achieves stereoscopic viewing by producing two panoramic images---one panorama for each eye.  We favored this technique as it does not require the composition of sub-images to recreate the projection effects~\cite{gledhill2003panoramic}.  

For our volume renderer, we modified its ray casting algorithm, such that the ray directions match those that are described in Google's ODS developer guide~\cite{ods}.  As suggested in the document, we used an interpupillary distance (IPD) of 6.4 cm, which was converted to match the units used by our renderer.  For advanced rendering techniques, we added pre-integration to alleviate sampling artifacts and volumetric shadows to improve depth cues.  We also incorporated the ability to change the clipping plane distance to prevent volume data features from being rendered uncomfortably close to the viewer.

\subsubsection{Timeline Editor}

Figure~\ref{fig:author_tool} (top-right) shows the timeline editor, which is a keyframe-based interface for the author to create a single instance of linear animation.  The timeline is made up of several independent lanes with icons to indicate the lane's respective dimension.  Each dimension can be changed over the animation sequence.  The author is able to preview their animation in the renderer and make any necessary edits.

Our authoring tool currently supports interpolation over camera, transfer function (TF), clipping planes, and temporal dimensions.  We briefly summarize the benefits of each dimensional change:

\begin{itemize}
\itemsep-0.10em 
\item{\textbf{Camera:} Viewpoint, or spatial, changes can help users have a better vantage point of the data set.  Camera changes include rotation, fly-through, and panning.  However, camera rotations--especially around the y-axis--may not be effective, since users already can view the content in any direction.  If possible, we recommend keeping the camera inside the data set, which will fully immerse the viewer in the content. }

\item{\textbf{TF:} Color mapping changes can isolate particular features of the data that fall on a specific range of values.  This can help the viewer focus on a certain feature, while the other characteristics are set to a lower opacity.}

\item{\textbf{Clipping planes:} Changing the positions of the XYZ planes--the planes that define an axis-aligned bounding box--can clip off values that fall outside the data's boundaries.  In some cases, moving clipping planes can reveal the internal structures of the data.  This can be particularly useful for medical data sets as it contains many internal structures for study.}

\item{\textbf{Temporal:} These changes are applicable to time-varying data sets, which show how the data evolves over the collected time steps.  Interesting data attributes may reveal themselves at certain time steps and not in others.  This dimension complements storytelling well, since it shows a natural progression of the data changing.}

\end{itemize}

In addition, the timeline editor includes features that ease the creation of animation, such as showing thumbnails of the data at the time of a dimensional change, color coding the keyframes to their respective dimension, and allowing the author to preview the changes in the renderer through scrubbing and playback of the timeline.

\subsubsection{Roadmap Interface}
The roadmap interface facilitates the creation of interactive, navigable videos.  Our approach centers around navigable video paths, which abstracts the filtering tasks--those required to visualize the meaningful attributes--from the end user.  These preset paths are represented in our roadmap structure, which has an underlying model and appearance of a node-link graph.  In our design, each node contains a single keyframe and each edge contains an instance of animation, which can be modified in the timeline editor.  An example roadmap is shown in Figure~\ref{fig:author_tool} (bottom-right).

To build navigable video content, the author must first build their roadmap. An author can either lay out the final roadmap structure, construct the edges systematically one-by-one, or have a workflow that is a mixture of the two methods.  Edges are visualized with an arrow from the source to target node which represent the starting and ending keyframes respectively.  Although video playback and dimension interpolation are bidirectional, the display of a directed edge provides authors a reference to which ends of the video they are modifying and connecting the dimensional changes.  In respect to the system, knowledge of a start and end of a video allows the interface to connect the animations seamlessly.

The resulting roadmap structure can be fairly arbitrary due to the interface's support for free-form creation.  For example, the length of the edge does not represent its animation's length.  However, the links between the nodes themselves determine how keyframes are shared across adjacent nodes.  These roadmap operations can be generalized to three categories:

\begin{itemize}
\itemsep-0.10em 
\item{\textbf{Build:} Whether along a single or already concatenated edge, content can be built upon in this linear fashion.}
\item{\textbf{Branch:} Video content can branch off from a mutual node.  This operation presents multiple video options for the viewer to choose.}
\item{\textbf{Merge:} Video content can be merged due to an earlier branch operation.  Keyframes are shared based on the connection order of the incoming edges to this mutual node.}
\end{itemize}

With the programmatic support of sharing keyframes, edges can be properly connected to ensure continuous animation amongst the adjacent video segments.

\subsection{Video Output}
\label{subsection:video_output}
\begin{figure}[b]
\qquad
\subfigure[Metadata file]
{
\centering
\theverbbox
}
\qquad
\subfigure[Roadmap]
{
\centering
\includegraphics[height=1.15in]{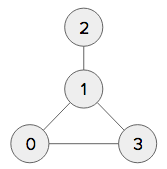}
}
\caption{A video output example. a) A simple metdata file that lists the video edges and connections with its end nodes. b) The corresponding roadmap from the metadata file.}
\label{fig:metadata}
\end{figure}

The authoring tool exports the roadmap along with several videos, where each video is associated with an edge.  The videos are encoded using the H.264 codec with FFMPEG.  This output is represented in a roadmap metadata file, which is used by the navigation stage to reconstruct the roadmap's connectivity. A video output example is shown in Figure~\ref{fig:metadata}.
\begin{verbbox}
Connectivity: 0 , 1
roadmap_0
Connectivity: 1, 2
roadmap_1
Connectivity: 1, 3
roadmap_2
Connectivity: 3, 0
roadmap_3
\end{verbbox}

Since our approach reduces the data exploration space, we wanted to support forward and backward video playback.  This allows viewers to traverse the videos at their own pace and review anything they may have missed.  Since video codecs do not support backwards playback, we generate two videos for both directions to implement this design feature.  Since stereoscopy is essential to include for enhanced depth and spatial cues, we also must generate two videos for the left and right eye to achieve parallax.  As a result, we have $4 \cdot n$ videos, where $n$ represents the number of edges in the roadmap.

As a roadmap grows and becomes more complex, the resulting memory footprint can grow significantly for a single viewing experience.  We mitigate the negative effects of this trend by choosing an appropriate Group of Picture (GOP) length. This value dictates how often a keyframe, or an \textit{uncompressed} frame, will be stored in the video file.  In addition, it affects seeking accuracy and memory size.  Decreasing the GOP length improves seeking accuracy while increasing the file's size.  Ideally, we want to have high seeking accuracy and a low memory footprint.  We use 0.25 seconds for the GOP length and found that it represents a good trade-off between these two factors.

We have encoded the video files with a frame rate of 30 frames per second (FPS), which is generally recommended for our target set of HMDs~\cite{purplepillVR}. We have found that this frame rate has a good balance amongst file size, I/O bandwidth, and latency. The frame rate of the playback interface is designed to stabilize around 60 FPS.

Since we are designing our system for VR headsets that require smartphones, we must be mindful of the available GPU resources, specifically, the number of video decoders.  We experimented with video resolution sizes that allowed four videos to be decoded for a given edge.  With our small benchmarking tool, we have found that 720p, or 1280x720 pixels, to be the maximum resolution for 360$^{\circ}$ videos that is supported by the mobile device's hardware.  Since this is a fairly low resolution, we supersampled the frames which were rendered as 4K, or 3840x2160 pixels, images to counterbalance visual artifacts such as aliasing.

\subsection{Navigational Interface}
Our system's navigational interface is the front-end component that presents the authored content to the user.  Between the authoring and navigation stages, the roadmap structure is preserved to determine how a viewer can traverse the content.  The roadmap is also presented to the viewer for reference on their progress within the authored content.  In the eyes of the viewer, each of the roadmap's edge represents a video and a node represents a position at either the start or end of the video segment.  A viewer is on an edge when viewing the video and is at an intersection when they reach either end of the video.  We have designed this interface to be simple and effective when guiding the viewer through the videos.  With its functionality similar to a virtual tour, our interface enables viewers to explore what the author has intended them to see.

\subsubsection{Head-Mounted Displays}
Nowadays, many HMDs are available to general consumers, such as Google Cardboard, Samsung GearVR, Sony PlayStationVR, Oculus Rift, and HTC Vive.  In contrast to specialized display systems like CAVEs, these headsets provide an affordable alternative to VR.  In particular, Google Cardboard is an accessible platform since it has been designed to be paired with an inexpensive viewing device and a smartphone. Naturally, the viewing quality is not as vivid compared to higher-end devices like the Oculus Rift or HTC Vive.  All of these HMDs feature head tracking, stereoscopic viewing, and at least one input element.  For this paper, we designed the navigational interface around the Google Cardboard platform, as it is the most affordable in the market.  

\begin{figure}[h]
 \centering
 \includegraphics[width=0.75\linewidth]{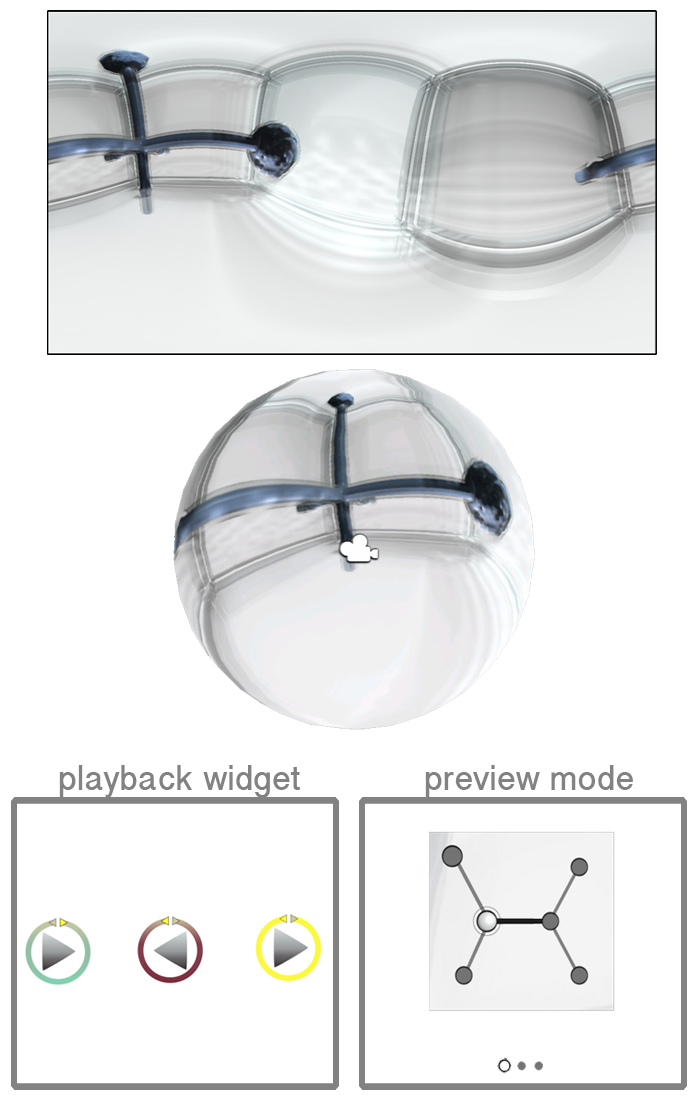} 
 \caption{An illustrative example of our Unity scene setup for our navigational interface. A panoramic video frame (top) is projected onto the sphere (middle) with a camera rig at its center.  A plane displays UI elements (bottom), either of one of the three playback widgets or preview mode.}
\label{fig:scene_setup}
 \end{figure}

\subsubsection{Interface Setup}
\label{subsection:playback_setup}
For development, we used Unity as our engine.  Unity offers cross-platform support for desktop and mobile deployment, along with native VR support and compatibility with popular HMDs.  As a result, we were able to port our interface to our target set of HMDs with little worry about device-specific development.  Since Unity currently does not provide support for video textures on mobile platforms, we used a third-party plugin to communicate with the video decoder for rendering the frames to our specified texture.

As shown in Figure~\ref{fig:scene_setup}, we have one scene set up to load the video content.  The video frames are mapped onto the sphere with a camera rig that is placed at the sphere's center. The camera rig contains two cameras that are offset by an 6.4cm IPD-equivalent in Unity's world coordinates.  To generate correct parallax, a camera either renders for the left or right eye.  Our user interface widgets are drawn on the UI plane in world space---versus screen space---to leverage the 3D stereoscopic effects from our scene setup.

To meet our design specifications, we have four active videos per roadmap edge during the viewer's experience.  Only two of the four videos are played at any given time to maximize GPU resources.  Based on the current playing direction, the appropriate two videos are queued to play, whereas the other two are paused.  When the viewer wants to change the playing direction or is at an intersection, we must switch and sync the pair of videos through seek operations.

\subsubsection{User Interface Design} 
\label{sssec:ui_design}
\begin{table}[b]
\centering
\begin{tabular}{c|c|c}
User state     & Button action & Playback action                                                       \\ \hline
On an edge         & Double tap    & Switch play direction                                           \\ \cline{2-3} 
                   & Tap + hold    & \specialcell{Play video \\ (release will pause video)} \\ \hline
At an intersection & Double tap    & \specialcell{Switch play direction \\ (back on edge)}                     \\ \cline{2-3} 
                   & Tap + hold    & Enter preview mode                                                   \\ \hline
                   & Tap           & Cycle video selection                                                 \\ \cline{2-3} 
Preview mode       & Double tap    & Exit preview mode                                                     \\ \cline{2-3} 
                   & Tap + hold    & Move onto selected video                                             
\end{tabular}
\caption{Available interactions in the navigational interface. Depending on the current playback state, the three recognized button actions will perform different playback actions. 
}
\label{tab:button_states}
\end{table}

We believe that an unobtrusive user interface will enhance the viewer's exploration experience.  Influenced by Oculus Connect 2's developer conference talk~\cite{oculusUIUX}, our interface utilizes a single button, in which we have defined the following actions: \textit{tap}, \textit{double tap}, and \textit{tap+hold.} The interface has three states, which is determined by the user's state: on an edge, at an intersection, or in preview mode.

\begin{table*}[htp]
\begin{center}
\begin{tabular}{|c|c|c|c|c|c|c|c|} \hline
 Data set  & Voxel size  & Total memory & \specialcell{Video \\paths}  & \specialcell{Memory\\ footprint}  &  \specialcell{Video\\length} & Avg. FPS & \specialcell{Dimensional\\changes}  \\ \hline
 Server room & 417x345x60  &  0.032 GB  & 5 & 0.13 GB & 00:01:47 & 60.401 & Camera \\ \hline
  \specialcell{Visible Human\\ (Male, Female)}  & \specialcell{Male: 512x512x1877\\ Female: 512x512x1734} &  \specialcell{Male: 1.83 GB\\Female: 1.69 GB} & 7  & 0.092 GB & 00:02:10 & 59.711 & \specialcell{Camera, TF, \\Clipping Plane}  \\ \hline
 \specialcell{Supernova\\(50 time steps)}  & 867x867x867 & 120 GB  & 1 & 0.54 GB & 00:01:40 & 59.943 & Temporal 
 \\ \hline
\end{tabular}
\end{center}
\caption{A quantitative summary of the case studies. Each data set is a floating-point volumetric field. Each field has a fixed size (Voxel size) for a single time step; only Supernova has multiple time steps. The amount of storage used by the raw data is listed under Total memory. The number of authored videos (Video paths), total file size (Memory footprint), and length (Video length) are shown next. The average frame rate (Avg. FPS) obtained in the case studies is also reported. Finally, we list the dimensional changes that were applied onto the data set. }
\label{tab:case_studies}
\end{table*}

We designed a playback widget that allows the viewer to see their progress on the current video.  The playback widget is structured as a circular progress bar with a play icon at its center.  At the start of a video, the progress bar is fully maroon.  When the user progresses forward, the bar fills in a counter-clockwise fashion with turquoise.  If the user plays the video backwards, the progress bar recedes and the play icon updates its direction.  At both ends of the progress bar are smaller triangles which are highlighted when the user has reached an intersection. With a \textit{tap+hold} action, the progress bar fills with yellow and only switches to the preview mode once the progress bar is full.  The preview mode enlarges, showing the roadmap to the user.  The nodes and edges have visual encodings which is determined by the user's viewing history.  In order to select a video, the user must \textit{tap} to cycle through the adjacent edges of their current node.  Below the roadmap is a set of dots, which represent the number of video options and the current selection.  A summary of the available interactions is available in Table~\ref{tab:button_states}.  

With the current design, users have no indication as to what the next path entails.  In our initial implementation of the preview mode, users were only able to view and choose from the upcoming changes.  This preview displayed one of four small multiples through a semi-transparent black frame.  The frame acted as a viewing window.  Each multiple displayed a video frame that corresponded to one of the four ``time steps," which were placed at 25 percent increments throughout the video.  In order to preview the video, we used alpha-blending transitions to cycle through the multiples.  However, when we conducted a pilot study prior to the usability study, participants reported this feature to be too confusing.  We decided to omit this preview feature in favor of showing the roadmap structure itself, which allowed users to be more aware of their global position in the authored content.  After more thorough design, this preview feature should be integrated back into the system and should be used in tandem with the roadmap.  This will reduce any randomness involved when users are deciding which path to traverse.

\section{Case Studies}
\label{section:case_studies}
We created a set of case studies to demonstrate an author's thought process when creating a presentation and how the resulting visuals present unique data features for the end user to learn.  These case studies include scientific data sets that vary in size and respective scientific domain. Table~\ref{tab:case_studies} summarizes the specifications of the data sets, along with a quantitative overview of the resulting video content.  The case studies were conducted on a Samsung Galaxy S6 that has 32 GB of storage, 3 GB of RAM, and a display resolution of 2560x1440 pixels.  For the following content, we used a renderer with advanced lighting features, which improve the depth perception of the data's features~\cite{lindemann2011influence}.  

\subsection{Server Room}
\label{ssec:server_room}
The server room data set is artificially-made and captures the characteristics of air pressure fields in a room full of machines.  With several rows of server machines, the room is expected to be hot which is damaging to computers.  To better maintain the machines, the owner can use visualization to evaluate the quality of their ventilation systems, which help regulate the room's temperature.  To visually present the characteristics of the room's air pressure, we used a heat map to not only color the level of air pressure, but also indicate the temperature at any point of the room; this shows where the ventilation could be improved.

The server room is the smallest data set of our three case studies, sized at 417x345x60 voxels and 0.032 GB.  For this case study, we used a single TF, which includes a rainbow color mapping with low opacity for the categorical representation of air pressure values---red and purple map to low and high air pressure, respectively.  This spectrum of warm to cool color hues maps to hot to cold temperatures.  The room and machines were colored gray to provide a contrast against the colors of the air pressure values.  We manipulated the camera dimension to provide a first-person view of someone walking through the room.  The authored paths extend from each corner and meet at the room's center.  We also created a path between two of the corners, as that particular hallway shows unique instances of air pressure distribution. 

\begin{figure}[t]
\centering
\includegraphics[width=1\linewidth]{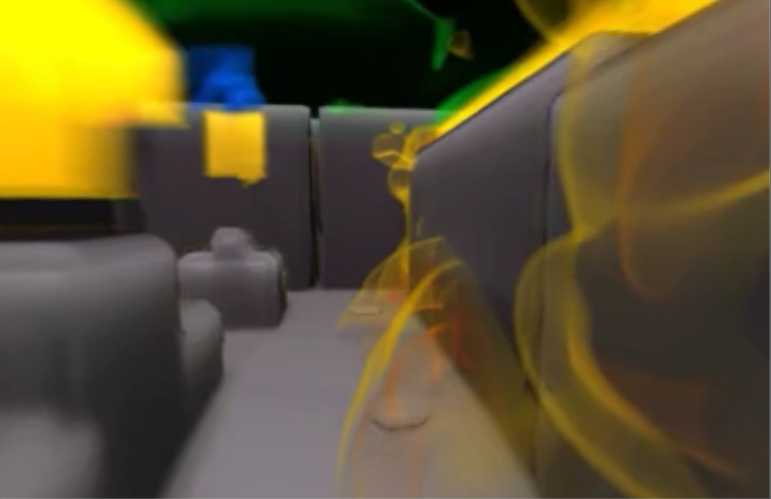}
\caption{A preview of interesting characteristics of the room's pressure field: blocks of medium air pressure hovering above the machines and floor vents with low air pressure.}
\label{fig:room}
\end{figure}

By moving the camera throughout the room, we allowed users to compare instances of air pressure from various parts of the room.  With the 360$^{\circ}$ viewing, users have control over where to examine the air flow and can determine whether the air is being emitted from floor vents or exhausted towards the ceiling.  Figure~\ref{fig:room} highlights a few interesting aspects from our presentation of the data, such as how low air pressure radiates from several floor vents.  Our TF also revealed several fairly-defined yellow blocks, which are shaped as the machines below them.  The color indicates that the air emitted from the machine's exhaust fan, which is not as powerful as the ventilation system.  

\subsection{Visible Human}

\begin{figure}[h]
{
\centering
\subfigure[Visible male]
{
\centering
\includegraphics[height=2.85in]{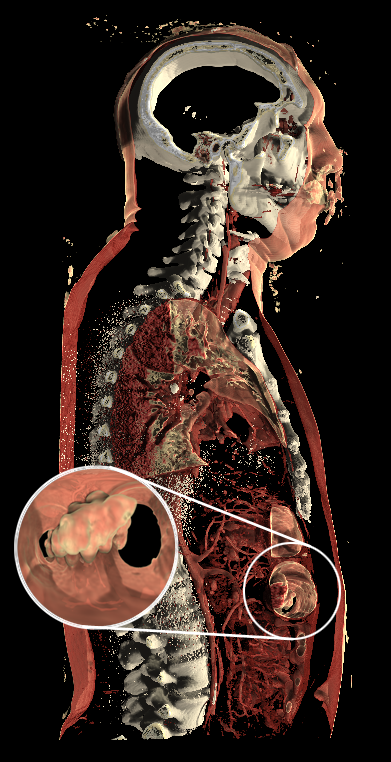}
}
\subfigure[Visible female]
{
\centering
\includegraphics[height=2.85in]{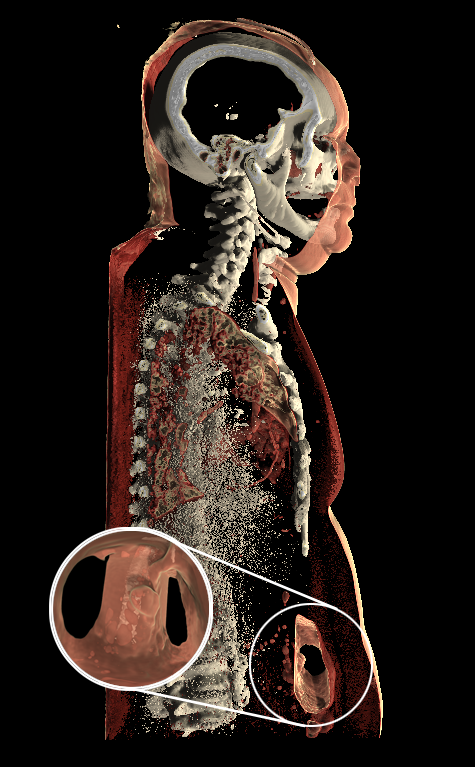} 
}
}
\caption{Side-by-side comparison of two sliced side views of the anatomy: a) male and b) female.  For instance, one navigable path goes through these lower body cavities in which the visual features are different; the annotation shows a panoramic view of inside the cavity.}
\label{fig:humanProject}
\end{figure}

The Visible Human data set is a collection of digitized slices of two full-body cadavers: one male, the other female.  This data set is provided by the U.S. National Library of Medicine's Visible Human Project~\cite{ackerman1998visible}, an effort that has captured high-quality cross-sectional photographs for visualizing the human body. Quality data such as this has opened more opportunities for study of the human anatomy.  We authored this data set to showcase the anatomical structures and provide a point of comparison between female and male bodies.

The male data set is 512x512x1877 voxels, whereas the female data set is 512x512x1734 voxels.  We first created content of the male by creating a roadmap with an edge that branched out with two options.  On the opposite end of the branch, we built another edge, where the video starts by fading-in the male from black.  For the other edges, we changed the following: moved the camera through two cavities in the head and lower chest, adjusted the clipping planes to reveal the internal structures, and fine-tuned TFs to filter out noisy values that was found in one of the explored cavities of one data set and not the other.  This same roadmap was used for the female, with a few modifications to correctly apply the dimensional changes onto the differing physical characteristics.  The two roadmaps generated two sets of content, which were connected by post-concatenating the videos that transitioned from black to its respective cadaver.  The metadata file was modified to reflect the concatenation of the male and female videos.

In the navigable content, we wanted to establish context by setting the camera outside each of the cadavers.  As expected, there are noticeable physical differences between the female and male.  By manipulating the TF to make skin values translucent and moving the slicing planes, we revealed the organs and bone.  As we moved the camera towards the cavities in the head or chest, users are able to observe differences at microscopic scales.  One instance is shown in Figure~\ref{fig:humanProject} (a, b), where the camera animates from the sliced view and into the lower chest cavity.  In the close-up, the male has a protrusion, whereas the female does not.  Overall, the nuances shared between the cadavers provided opportunities for the user to study. 
 
\subsection{Supernova}
The supernova data set was created from the results of physical model simulations on a supernova star.  The value visualized is entropy, or the rate of decline in energy.  These simulations tend to be large, complex, and multi-modal, which can be difficult for scientists to quantitatively analyze, let alone users like astronomy students.  When authoring these videos, we wanted to simplify the experience to the evolutionary changes of a supernova's energy.  

\begin{figure}[t]
\centering
\subfigure[Time step 1]
{
\includegraphics[width=1\linewidth]{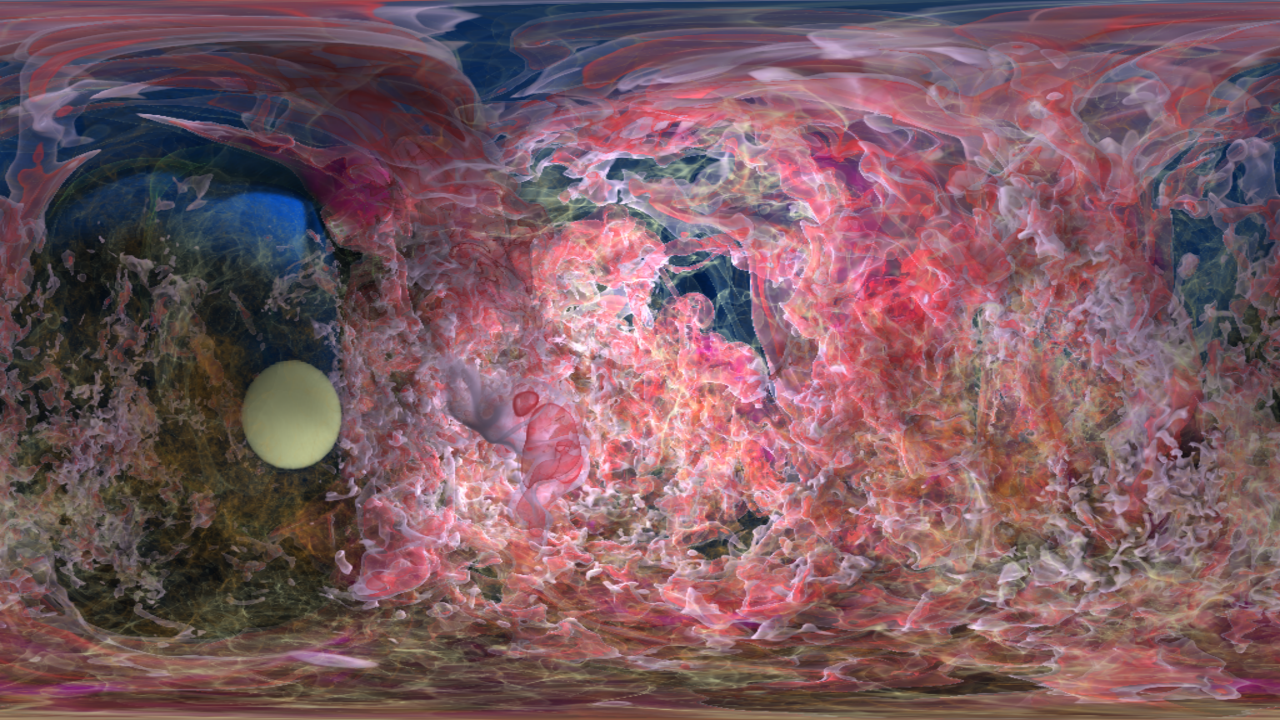} 
}
\\
\subfigure[Time step 50]
{
\includegraphics[width=1\linewidth]{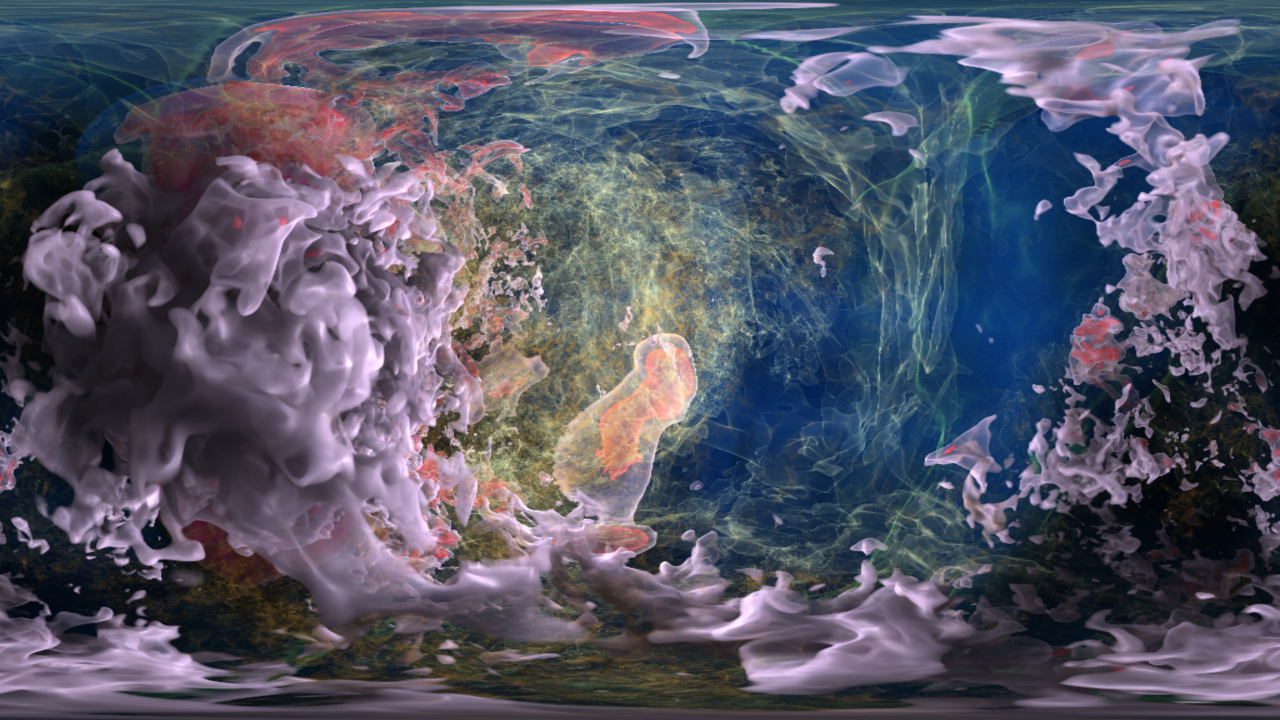}
}
\caption{Panoramic views of the supernova  at different time steps: a) 1 and b) 50.  The gases have experienced changes in movement and energy.}
\label{fig:supernova}
\end{figure}

In this case study, we visualized 50 time steps.  The data set is sized at 867x867x867 voxels, in which a single time step is 2.40 GB.  We used a single TF for coloring entropy---blue is low and purple is high.  By restricting the use of a single TF throughout the videos, we maintained a consistent visual encoding between color and energy.  In order to fully-immerse the viewer, we fixed the camera's position at the center of the supernova and near its core.  Since the total memory to store all the time steps amounts to 120 GB, we generated the content in segments--10 time steps at a time.  Then, we concatenated the segments for viewing as a single video.

Figure~\ref{fig:supernova} (a, b) shows the supernova's dynamic nature over the 50 time steps. We chose to show a large range of time, so users can see how the star's gases evolved in terms of movement and energy: The gas clouds have wrapped around the core and seem to have experienced high entropy, which can be inferred by its color transition to purple.  Also, the resulting memory footprint is 0.54 GB in contrast to the original data size of 120 GB, which is an impractical size for average computers to hold in memory and interactively render for visual presentation.  With results like this, our system makes large data sets more accessible for presentation in a modest setting.

\section{Formative Usability Study}
\label{section:usability}
We conducted a small usability study to assess how well-received the presentation experience is for a student audience.  This study serves as a formative evaluation, which will help us ascertain the strengths and weaknesses of the interface.  Since the interface allows users to dictate the playing and viewing direction of the navigable videos, it was important that we receive feedback on its current design.

By assuming the role as author, we created a presentation to highlight the data set's unique features.  We chose to simplify the viewing experience by only introducing camera movement as the only data dimension that users can change as they play the videos.  We asked participants to answer four questions to encourage them to navigate through the room to find the answers.  The task was intended for the students to interact and use our presentation system, such that the correctness of their answers hold little significance to our study's results.
 
 \subsection{Procedure} 
First, we summarized the goals of our research and pre-assessed the user's experience on the concepts involved in our system.  Shortly after, we had the participants go through a tutorial that walked them through the operations to navigate through a sample data set.  For those who were not familiar with Google Cardboard, we reviewed the headset's features.  The tutorial also covered how to use the preview mode, which introduced the roadmap.  It explained that each edge represents a video and how at each path's end, users would be presented options on where to go next in the roadmap.

Secondly, we asked the users to complete a navigation task as a simple exercise in using the navigational interface.  Users were not required to answer the questions in a particular order and were not timed per answer.  This task involved the same navigable videos of the server room data set, which is described in Section~\ref{ssec:server_room}.  These questions were designed to be answered objectively and solved by leveraging the 360$^{\circ}$ viewing and moving throughout the room.  This task required them to answer the following questions:
 
\begin{itemize}
\itemsep-0.10em 
\item{In any two room’s corners, is the pressure high or low?}
\item{Are the exhaust vents on the top or sides of the machine?}
\item{Which color (s) emit from the floor vents?}
\item{Is high air pressure on the floor or on top of the machines?}
 \end{itemize} 

A post-assessment followed once the user finished answering the questions.  They were asked to rate their thoughts against a series of statements on a 5-point Likert scale.  Finally, we asked if they had any feedback for improving the playback interface.
 
\subsection{Participants}
Using the university's emailing-lists, we recruited 22 students who are currently involved in the STEM fields, such as Computer Science, Biomedical Engineering, Physics, and Material Sciences.  16 were male and six were female.  Participants' mean age was 25.  For the pre-assessment, students reported an average rating of 3.18 ($\sigma$=0.89), 3.14 ($\sigma$=1.04), and 3.18 ($\sigma$=1.05) for familiarity with scientific visualization, VR, and 360$^{\circ}$ videos, respectively.  In our study, we had two participants who could not complete the navigation task due to the phone overheating and having experienced high levels of cybersickness.
 
\subsection{Environment}

Participants were asked to sit at a table in a swivel chair.  By sitting in a swivel chair, the user can better align their body when viewing in 360$^{\circ}$.  In front of them were reference sheets about the interface widgets and data set, along with a copy of the questions.  Users could freely refer to these materials at any time during the study.  The VR devices used were a Google Cardboard headset and the same phone described in Section~\ref{section:case_studies}.  A Google Chromecast streamed the phone's screen to a secondary monitor for us to troubleshoot any issues users ran into.  Only audio was recorded for user's feedback which were reviewed after the session.  Users were encouraged to take as much time as they needed when exploring the content and answering the questions. 
\subsection{User Feedback}
During the post-assessment, we received valuable user feedback on the navigational interface.  Most subjects wanted the roadmap to be displayed at all times for reference on their location.  Some subjects suggested new usability features.  S3 suggests \textit{``maybe if you triple tap [it] show[s] help.''}  This help menu would display content similar to that of Table~\ref{tab:button_states}.  Others related the navigational interface back to their studies.  S10 explained \textit{``I just know this thing in 2D---pressure dispersion around a room. 2D would be as useful to me as 3D.  If I didn't know that already, this would be much more useful as a learning experience.''}  S5 stated \textit{``I look at proteins on my computer, and so it is very annoying to look on a desktop and just drag it around with a mouse, so I was thinking how nice it is to have 3D.''} 

Although in a controlled setting with a fairly simple data set, participants had fair reviews of our navigational interface.  Students reported an average rating of 3.41 ($\sigma$=0.85), 3.86 ($\sigma$=0.77), and 3.67 ($\sigma$=0.86), for interface usability, presentation effectiveness, and if they would use the interface again, respectively.  However, we did observe that a fairly high learning curve existed when using our system.  For example, we noticed users showing little aptitude when viewing 360$^{\circ}$ videos, which may have negatively influenced their experience.  Specifically, some users were disoriented when the camera movements did not align with their current viewing direction.  We noticed these participants passively viewed the content and did not leveraging the 360$^{\circ}$  viewing.  Overall, most participants still expressed excitement and piqued interest towards the future applications that our system enables.

\section{Discussion}
\label{chapter:discussion}
When considering all the possible video configurations, we expected that videos of long length would be frustrating and uninspiring for viewers.  Originally, the experience was intended to be ``continuous," such that it emulates real-world experiences of navigating through an area.  However, this interactive tour can easily become a maze, where the user becomes lost or impatient.  For example, S8 expressed that they wanted a teleportation feature.  They expressed frustration when traversing the video back again to reach an already-known area of interest.  However, we observed that the participants made a strong connection of the roadmap's layout to spatial movement. More design consideration will be needed when introducing abstract dimensional changes, such as TF, to the user and breaking this seemingly strong connection between spatial movement and the user's change in position on the roadmap.  It is unclear if the ``convenience" of teleporting outweighs the likely jarring effect when viewing the possibly large and discrete visual changes.

The participants' reception on the authored content highlights the tight interplay between the authoring tool and navigational experience.  For instance, long videos are likely to be tedious for a participant to traverse through.  If the path happens to be a ``dead end," the user must play the video backwards to reach another intersection for video selection.  Also, since there is no system restriction on how the content can be authored, extreme cases of content--to name a few, a large number of short videos or one extremely long video--can negatively impact the playback experience.  With further experiments to cover all bases of roadmap configurations along with an evaluation involving the intended users, that is, author and viewer, we can build a reference of best practices when creating content for immersive navigation.  It is important to balance the user's experience with the ease and flexibility of authoring content--viewers should feel engaged and in control, while authors should be able to create whatever they desire to fit their presentation needs.

\section{Future work}
Our system's design is flexible and should be applied to other disciplines that would benefit from visualization, animation, and the 3D space that VR offers.  Information visualization is traditionally rendered in 2D and often involves extremely large amounts of data in which the relationships are often aggregated or filtered due to the data's density and lack of rendering and viewing real-estate~\cite{schulz2013grooming}.  For these reasons, immersive, stereoscopic viewing may be a viable solution to these challenges with its extra spatial dimension to lay out the data.  For this class of renderers to adopt our approach, a set of dimensional changes must be specified.  Some of these dimensions include temporal, motion, and color and filtering transitions.

\subsubsection{Authoring Tool}
We have found that 360$^{\circ}$ viewing has its pros and cons: Users have full control of where they are looking, but are susceptible to missing crucial aspects of the content.  Although in favor of an interactive environment, these drawbacks are detrimental to how well the author's intent is conveyed when viewing the tailored experiences.  One feature that may help is incorporating text annotations into our system.  These annotations would be fairly arbitrary, ranging from displaying instructions to interesting facts about the data set.  Using the authoring tool, the scientist can specify their message and where the annotation resides.  The annotations will then exist on the playback side for the viewer to encounter them.  We can experiment with how to present these annotations--for example, as pop-ups or ``signs"--to see what best aids the viewer to understand and learn the presented content.  We want the annotations to serve as guides to ensure that the viewer still has sufficient interactivity and control of their viewing experience.  

We also would like to have comprehensive previewing features to aid scientists in creating more effective, interactive narratives of their data and findings.  For example, authors should be able to traverse through the roadmap to preview what their tour would look like.  Another example is to provide a complementary rendering window which would emulate the viewer's experience, especially when the content is projected onto the sphere. It also may be possible to port a frame or animation segment directly to the physical HMD.  This allows the scientist to experience the spatial depth that stereoscopic effects provide.  However, exporting the content between the desktop and HMDs in a streamlined manner is not trivial.  

Another step for this work is to evaluate our authoring tool when used by scientists.  By integrating our tool into their domain-specific workflows, scientists can present their findings in our new medium.  During the evaluation, it is important to observe how scientists use the tool.  Depending on the sample size, we may find a trend in how scientists use the tool which dictates the different constraints our system should enforce.  The scientists' feedback would be insightful, allowing us to better align the software design to meet their needs.  In practice, most scientific animations are created by skilled animators.  Since scientists are unlikely to be experts with animation tools, the feedback can guide later designs, whether it involves an interface redesign or more scientist-friendly features.  

\subsubsection{Navigational Interface}
\label{section:discussion_video}
In future work, we would like to improve our video memory footprint for each viewing experience.  Since we used a third-party plugin to render video frames to a texture, we had less flexibility for memory optimizations.  In the next iteration, we would like to implement our own video encoder and decoder, which we can design to fit our needs.  We can employ techniques such as Facebook's pyramid encoding~\cite{pyramidFB} and benefit from its data management scheme, which has reported to reduce memory up to 85 percent.

A major challenge in immersive viewing is prompting what a user should look at.  Since the 360$^{\circ}$ viewing is fully interactive, a user decides what they end up looking at, and possibly misses salient information.  We believe that on-screen clues, such as the aforementioned annotations from the authoring stage, or audio cues can address this issue.  The on-screen clues can indicate where the user should turn their head, whereas immersive audio can be another perceptual channel to prompt users where to look.  Audio can be particularly useful for learning and captions can also be used to increase accessibility.  However, the bidirectional video playback presents a possible challenge since audio and reading is naturally linear.  Smaller, ``discrete" audio cues would be easiest to use.  More design and experimentation would be required to integrate audio into our system.

Following through with the nature of storytelling, we would like to facilitate the sharing of this scientific content.  A centralized repository can help form a community that builds and shares these immersive, visual presentations of scientific data.  In addition, real-time streaming of the authored videos enables our system to support a multi-user experience.  However, we must be wary of the consequences: Network dependencies are likely to be introduced in order to enable content distribution.  Although we can offload the number of videos resident on the device's memory and GPU decoders, we would have to fine-tune the video settings to be better suited for streaming video---if latency is poor, it will make for an uncomfortable viewing experience.

\section{Conclusion}
Many scientific studies are about capturing and understanding complex physical phenomena and structures. Immersive visualization offers a more perceptually effective way to examine 3D structures and spatial relationships.  Capturing this immersive space, our presentation medium leverages a scientist's expertise to display the content effectively.  For viewing, the navigational interface is compatible with increasingly affordable HMDs, which are accessible VR platforms for showcasing a scientist's research to their target audience.  From the case studies and formative usability study, our findings suggest that our navigable videos show promise as a presentation medium.  However, the interface will need further design iterations to improve its usability, especially with the expectations that it will be used by people with varying interest levels in scientific visualization.  We believe that our work can be used in other visualization fields, such as information visualization and visual analytics, which would benefit from an immersive presentation medium.

\acknowledgements{
This research was sponsored in part by the UC Davis RISE program, US National Science Foundation via grants DRL-1323214, IIS-1528203, and IIS-1320229, and U.S. Department of Energy via grant DE-FC02-12ER26072.}

\bibliographystyle{abbrv}
\bibliography{immersive_HMDs}

\end{document}